\documentclass[twoside,leqno,twocolumn]{article}

\usepackage[letterpaper]{geometry}

\usepackage{ltexpprt}

\usepackage{cite}
\usepackage{amsmath,amssymb,amsfonts}
\usepackage{graphicx}
\usepackage{textcomp}
\usepackage{xcolor}
\usepackage{hyperref}
\usepackage{todonotes}
\usepackage{algorithm}
\usepackage{algpseudocode}
\usepackage{tabularx}
\usepackage{textcomp}
\usepackage{chronology}

\def\scriptO{{{\it O}\kern -.42em {\it `}\kern + .20em}}
\def\RR{{{\rm l}\kern - .15em {\rm R} }}
\def\PP{{{\rm l}\kern - .15em {\rm P} }}
\def\L2{{{\sf L}^2}}
\def\H1{{{\sf H}^1}}
\def\PN2{{\PP_{N}-\PP_{N-2}}}

\def\complex{{{\rm C} \kern - .53em {\rm l} \kern + .38em}}

\def\a1{{ | \lambda_{\min} |}}

\def\l1{{   \lambda_{\min}  }}

\def\bu0{{\underline {\bf 0}}}

\def\bu{{\bf u}}

\def\ub{{\underline b}}

\def\ud{{\underline d}}

\def\uu{{\underline u}}
\def\uv{{\underline v}}
\def\uw{{\underline w}}
\def\ux{{\underline x}}

\def\uy{{\underline y}}
\def\uz{{\underline z}}
\def\u0{{\underline 0}}
\def\1u{{\underline 1}}

\def\bL{{\mathbf{L}}}

\def\BibTeX{{\rm B\kern-.05em{\sc i\kern-.025em b}\kern-.08em
    T\kern-.1667em\lower.7ex\hbox{E}\kern-.125emX}}

\begin{document}

\title{parRSB: Exascale Spectral Element Mesh Partitioning
\author{Thilina Ratnayaka\and Paul Fischer}
}

\maketitle

\begin{abstract}
  We introduce {\em parRSB} -- a parallel, highly scalable graph partitioner for spectral
  element {\em meshes} that produce high quality partitions.
  {\em parRSB} is based on Recursive Spectral Bisection (RSB) algorithm implemented on the
  dual graph of the input {\em mesh}.
  RSB uses the {\em Fiedler vector}, which is the eigenvector associated with the smallest
  non-zero eigenvalue of the Laplacian matrix of the dual graph for making partitioning
  decisions and tries to minimize the communication volume between the partitions.
  We implemented two numerical methods: {\em Lanczos}, and {\em Inverse iteration} using
  Conjugate Gradient method to compute the {\em Fiedler vector}.
  We present partitioning results using {\em parRSB} on Summit and Frontier supercomputers
  at Oak Ridge National Laboratory to illustrate the quality of the partitions produced
  by {\em parRSB} and the scalability of our implementation.
  We also present results for some of the optimizations we did to speed up the partitioning
  process.
\end{abstract}

\section{Introduction}\label{sec:intro}

High Performance Computing (HPC) landscape is changing rapidly from CPU based systems to accelerator
based systems powered by General Purpose Graphics Processing Units (GPGPUs or simply GPUs).
This is evident by the fact that 7 of the top 10 computers of the latest TOP500 list of
supercomputers~\cite{top500} are accelerator based systems.
The fastest, Frontier supercomputer at Oak Ridge Leadership Computing Facility (OLCF) using AMD
Instinct\textsuperscript{\texttrademark} MI250 GPUs is the first and the only exascale system on the
list so far.
A second exascale system, Aurora at Argonne Leadership Computing Facility (ALCF) is expected to be
deployed by the end of 2023 and will also be an accelerator based system with
Intel\textsuperscript{\textregistered} GPU Max series GPUs.
These exascale and other pre-exascale systems like Summit (also at OLCF) have delivered an unprecedented
level of peak performance and storage capacity thus enabling scientific community to solve larger and more
complex problems faster than ever.
Spectral Element Method (SEM) community has always been among the first to use the state of the art
supercomputers in leadership computing facilities for numerical solution of Partial Differential Equations (PDEs).
In this paper we are going to revisit an important aspect of high performance SEM simulations -- problem
decomposition or simply known as {\em partitioning} in the context of the changing HPC landscape in the
exascale era.

SEM simulations are performed on a {\em mesh}, which is a triangulation of the computational domain
using smaller regions called {\em elements}.
These elements usually have the shape of quadrilaterals in 2D and hexahedra in 3D.
Each {\em element} is further discretized into a set of points called {\em nodes} or degrees of
freedom (DOF).
Numerical (or approximate) solution is represented by a set of \textit{basis functions} defined on
these nodal points which provide a polynomial approximation to the solution of the underlying PDE.
Number of {\em elements}, $E$, and the order of {\em basis functions}, $N$, are selected based on the
geometric complexity of the computational domain and the accuracy requirements of the simulation respectively.
Size of the problem, $n$, measured in terms of the number of degrees of freedom (DOF) is equal to
$EN^d$ where $d$ (usually 2 or 3) is the physical dimension of the problem.
In this paper, our focus is on $d=3$ since it has the highest practical importance in science and
engineering applications.

\begin{figure}[htb]
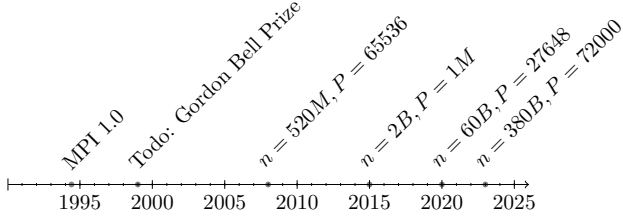

\begin{chronology}[5]{1990}{2025}{55ex}
\event{\decimaldate{1}{06}{1994}}{MPI 1.0}
\event{1999}{Todo: Gordon Bell Prize}
\event{2008}{$n=520M, P=65536$}
\event{2015}{$n=2B, P=1M$}
\event{2020}{$n=60B, P=27648$}
\event{2023}{$n=380B, P=72000$}
\label{fig:semfem_timeline}
\end{chronology}
\caption{Timeline of important events in high performance scientific computing and the evolution
of the problem size measured in degrees of freedom (DOF).}
\end{figure}

Figure~\ref{fig:semfem_timeline} shows a timeline of important events in high performance SEM simulations
and the evolution of the problem size, $n$.
Introduction of Message Passing Interface (MPI)~\cite{gls99} in 1994 was a major milestone in high
performance scientific computing since it standardized the use of distributed memory parallel computing
with explicit message passing.
MPI has since become the de facto standard for parallel computing and is used in almost all leadership
class computing facilities enabling users to utilize thousands of processors in parallel.
In 2008, a SEM simulation using $520$ million DOF was performed on $65536$ processors on BG/P machine at
ALCF~\cite{fischer08a} using Nek5000~\cite{nek5000} -- a highly scalable spectral element solver optimized
for the CPUs.
In 2015, another Nek5000 simulation had $2$ billion DOF on $1$ million processors on ALCF's Blue Gene/Q
machine~\cite{fischer2015scaling}.
Once the Summit computer at OLCF was commissioned with NVIDIA V100 GPUs, NekRS -- a GPU port of Nek5000
using OCCA~\cite{medina2014occa} library was used to perform a 60 billion DOF simulation using
$27648$ GPUs~\cite{fischer2022nekrs}.
Then in 2023, NekRS was used in a massive simulation with $380$ billion DOF on $72000$ GPUs on the first
exascale system -- Frontier at OLCF.
It is clear that the problem sizes will soon reach {\em Trillion} DOF on future exascale systems with
$10^9$ -- $10^{10}$ spectral {\em elements}.
Also, it is worth noting that the local problem size, $n/P$, is significantly larger on the GPU based
simulations compared to the CPU based simulations.

With problem sizes getting larger and larger, an important metric for a user of a HPC system is how
fast one can get to the solution i.e., {\em time-to-solution}.
In the context of parallel SEM simulations using $P$ processors, time spent on the local computation
by processor $p$, $T_{a,p}$ is proportional to the local problem size $n_p$ on that processor and
the arithmetic time $t_a$ -- which is usually measured as the time taken to perform a single floating
point operation (FLOP) in small matrix-matrix multiplications.
If $F$ FLOPs are performed per DOF by the numerical method to find the approximate solution, then the
arithmetic time is given by:
\begin{equation}
T_{a,p} = F\cdot n_p \cdot t_a
\label{eq:arithmetic_time}
\end{equation}
where $\sum_{p=1}^P n_p = n$.
When it comes to communication time, $T_{c,p}$, there are two main factors -- total message size (or
volume) $W_p$ measured in 64-bit words and the number of messages $M_p$.
Then using the postal model~\cite{fischer2015scaling}, the communication time can be approximated as:
\begin{equation}
T_{c,p} = \alpha M_{p} + \beta W_{p}
\label{eq:communication_time}
\end{equation}
where $\alpha$ is the latency in seconds and $\beta$ is the inverse bandwidth in seconds per word.
Total time is then given by the sum of the arithmetic and communication times:
\begin{equation}
T(n, P) = \max_{1\le p \le P}\left[T_{a,p}\right] + \max_{1\le p\le P}\left[T_{c,p}\right]
\label{eq:total_time}
\end{equation}
Based on~\ref{eq:total_time}, it is clear that we have to minimize both arithmetic and communication
time in order to reduce the total time.
Since the arithmetic time is proportional to the local problem size, the minimum of the maximum
arithmetic time is achieved when the problem is evenly distributed among the processors.
Thus, {\em load balance} is important to reduce the total time.

Minimizing the communication time is more complicated since it depends on the communication volume
and the number of messages.
The cost of the two terms in Equation~\ref{eq:communication_time} is equal when
$\alpha/\beta = W_p / M_p$.
It is easy to see $m_2 = \alpha/\beta$ is the size of message that would take as twice as long to
transmit as it would take to transmit a single word.
If the average message size, $W_p/M_p$ is less than $m_2$, then the communication time is
dominated by the number of messages, otherwise, it is dominated by the communication volume.
We will demonstrate that for the GPU based supercomputers, the average message size will be much greater
than $m_2$ more often than not and the communication time will be dominated by the communication volume.
This is due to the fact that GPU based systems usually have a much larger local problem size and thus
larger total message sizes.
On traditional CPU systems where the local problem was much lower, the communication time was dominated
by the number of messages.
So a partitioning method which reduces the {\em communication volume} is essential for exascale SEM
simulations.

%
The remainder of the paper is organized as follows.
We define the graph partitioning problem on a mesh in Section~\ref{sec:problem} and then look at the
related work in Section~\ref{sec:related} followed by a description of our approach
Sections~\ref{sec:rsb}--\ref{sec:inverse}.
We present results in Section~\ref{sec:results} followed by a discussion of future directions in
Section~\ref{sec:future}.

\section{Problem Definition}\label{sec:problem}

Since we want to partition {\em mesh elements} to processors, we consider the dual graph $G(V, E)$ of
the original {\em mesh}.
Here $V$ is the set of vertices of $G$ where each vertex represents a spectral element in the original
{\em mesh}.
$E$ is the set of edges in $G$ which represent the connections between spectral elements.
Edges exist between elements which share a vertex, edge, or face in the original {\em mesh}.

Given a graph $G(V,E)$ with non-negative edge weights $\omega : E \rightarrow \mathbb{R}_{>0}$ and $P$
processors, the graph partitioning problem as it relates to mesh partitioning is to find a set of subsets
$V_1, V_2, \dots V_P$, such that:
\begin{eqnarray} \label{eq:partition}
  V_1 \cup V_2 \cup \dots V_P &=& V \\
  V_i \cap V_j &=& \emptyset,\,\, i \ne j,\, 1 \le i, j \le P \\
  \max{\mid V_i \mid - \mid V_j \mid} &=& 1,\,\,1 \le i, j \le P
\end{eqnarray}
where $\mid V_i \mid$ is the size of the subset $V_i$.
The first two constraints ensure that subsets $V_i, \;1\le i\le P$ are a partition of $V$ and that the
subsets are disjoint (i.e., each {\em element} is assigned to exactly one processor).
Our weight function $\omega$ is the number of shared vertices between two {\em elements} connected by
the edge $e \in E$.
This could be $1$ if the two {\em elements} share a vertex, $2$ if they share an edge,
or $4$ if they share a face in the original {\em mesh}.

\section{Related Work}\label{sec:related}

Finding the optimal partition (either bisection or a $k$-way partition) is
known to be NP-complete. Optimal solutions are rarely obtained and
many heuristics have therefore been developed to address this problem,
which is important in many fields~\cite{schloegel2000graph}.
These heuristics try to group elements based on similarity measures like
spatial position of the mesh elements or how strongly the elements are connected.
The former set of methods are known as \textit{geometric methods} and the
latter are known as \textit{combinatorial methods}.  \textit{Spectral methods}
on the other hand solve an optimization problem that tries to minimize the
edge-cut between partitions using the global connectivity information of the
graph.

Geometric methods make the partitioning decision based on the physical
coordinates of the graph nodes in space ignoring the graph connectivity.  Such
methods are applicable only when nodes of the graph under consideration are
associated with a physical space or when the connectivity information can be
used to embed the graph nodes in a $k-$dimensional
space~\cite{schloegel2000graph}.  Since the SEM/FEM meshes represent a
computational domain in physical space, it is straightforward to apply these
methods.  Because connectivity information is ignored, geometric methods tend
to produce partitions with lower quality than combinatorial or spectral
partitioning methods.

Recursive Coordinate Bisection (RCB) seeks to reduce the communication by
trying to minimize the length of the boundary-cut between partitions.  To
minimize the boundary, RCB finds the longest coordinate axis and bisects the
graph normal to that axis.  This approach avoids cutting along the longest
axis.  RCB is extremely fast due to the fact that we only need a fast sorting
algorithm to bisect the graph in parallel.  Apart from giving low quality
partitions, RCB can only partition along the normal coordinate axes.  This can
affect the quality of partitions as the longest boundary may not be a normal
coordinate axis, for example when the mesh is oriented at an angle to
coordinate axes~\cite{schloegel2000graph}.  Recursive Inertial Bisection fixes
this issue by calculating the inertial axis of the mesh, projecting the element
coordinates to this axis and then ordering elements by this projected
coordinate.  Both of these schemes can be improved by rescaling the geometry
such that, on average, the elements are of the same diameter, which avoids
having skewing  the long axis by high-aspect-ratio elements
~\cite{heath1995cartesian,gilbert1998geometric,berger1987partitioning,schloegel2000graph}.

Both RCB and RIB are based on the projection of coordinate values
in a single axis.  Space-filling curves try to improve over RCB/RIB algorithms
by positioning center of mass of mesh elements in a continuous curve in higher
dimensions (2D, 3D).  After elements are placed in a curve, that ordering can
be used to partition into the required number of subdomains~\cite{pilkington1996dynamic}.

Combinatorial methods try to use the connectivity information to group highly
connected vertices, thus producing partitions with lower edge-cuts that tend to
have a lower number of disconnected components compared to geometric methods.
Levelized Nested Dissection (LND) starts with a single vertex and tries to find
a subdomain which contains half of the graph nodes where all the element of the
subdomain are connected.
LND achieves this objective by gradually growing subdomains through the
addition of adjacent vertices.  Although LND produces better partitions, it is
slower than geometric methods since the algorithm doesn't have the same level
of parallelism as the previous approaches~\cite{george1981computer}.

Spectral methods try to solve a relaxed version of a discrete optimization
problem.  Let $G(V,E)$ be a graph with $\mid V \mid$ vertices.  Graph
bisection can be thought of as splitting $\mid V\mid$ vertices into roughly equal sets
$A$ and $B$ in size while minimizing edge-cuts.  Let's consider a vector
$\underline{x}$ of size $n$ where each entry in $\underline{x}$ is either $+1$
if $i^{th}$ vertex of graph belongs to $A$ or $-1$ if it belongs to $B$.  If
$\mathbf{L}$ is the unweighted Laplacian associated with the graph (which we
will define in Section~\ref{sec:rsb}), then
$\underline{x}^T\mathbf{L}\underline{x}$ is equal to $4\delta(A, B)$ where
$\delta(A, B)$ is the number of edges shared between $A$ and
$B$~\cite{pothen1997graph}.  Now minimizing edge-cuts is equal to minimizing
$\underline{x}^T\mathbf{L}\underline{x}$ subject to the constraint $x_i = \pm
1$.  This minimization problem itself is NP-Complete and can't be expected to
be solved exactly.  The problem can be relaxed by allowing each component $x_i$
to vary continuously and the minimizer for the relaxed problem is the
eigenvector corresponding to the smallest positive eigenvalue of
$\mathbf{L}$~\cite{pothen1997graph}.

Parallel implementations of these algorithms are available in several software
packages.  parMetis is a Message Passing Interface (MPI) based parallel library
that implements a variety of algorithms for partitioning unstructured
graphs~\cite{karypis1997parmetis}.
The main algorithm in parMetis is a multilevel $k$-way partition algorithm
which works by first coarsening the input graph to a smaller graph with a few
hundred vertices and then projecting the computed $k$-way partition for this
smaller graph to the original graph.  PT-Scotch also uses a similar multilevel
approach but with nested dissection at the top level to partition the projected
separator set rather than $k$-way partitioning~\cite{chevalier2008pt}.  They
also use additional band refinement heuristics to refine the identified
projected separators.  Other notable software packages include
JOSTLE~\cite{walshaw2007jostle} which also performs multilevel partitioning and
Zoltan~\cite{devine2006parallel} which performs hypergraph partitioning.

While there is apparently a wealth of partitioners available, not all of them provide the level of control
and scalability required for targeted exascale simulations using the SEM.
Because of the SEM granularity, where each element typically has 500 to 2000 degrees-of-freedom, we need
strict load balance.
Furthermore, the partitioner must scale to $P=10^6$--$10^7$ ranks for current and future platforms such as
Frontier, Aurora, etc.
Users of the SEM, generally engineers and physicists, need an end-to-end solution where they provide minimal
data input (e.g., a mesh); derived quantities such as partitions should be dealt with automatically.
The partitioner must support multi-material domains, for example, such as in conjugate heat transfer where
there is an expensive (i.e., flow) region that dominates the work per element, coupled to a typically less
expensive region where one only needs to solve for a scalar field (temperature).
Partitioning both these regions to achieve reasonable {\em load balance} and {\em minimal communication}
requires detailed control over the partitioning process.

There are multiple reasons, therefore, to revisit the parallel partitioning question on modern architectures.
In this paper we introduce {\em parRSB}, a parallel graph partitioner based on Recursive Spectral
Bisection (RSB) described in Section~\ref{sec:rsb} to achieve {\em load balance}, {\em minimal communication volume}
and detailed control over the whole partitioning process.
Our main contributions include space-efficient storage and evaluation of the Laplacian based on a scalable
gather-scatter operator, augmented preconditioned projection for the inverse iteration, approximate
Krylov-subspace projection of the inverse iterates, algebraic multigrid preconditioner (AMG) for solving the
Laplacian system and use of RCB as a pre-partitioner to speed up the partitioning process.

\section{Recursive Spectral Bisection}\label{sec:rsb}

The basis of Recursive Spectral Bisection (RSB) is a graph bisection algorithm
first developed by Pothen {\em et al.} \cite{pothen1990partitioning}.
The main idea of their work is to make the bisection decision based on the
eigenvectors of the Laplacian matrix $\mathbf{L}$ of the graph $G(V, E)$ where
$G$ is the dual graph defined in Section~\ref{sec:intro}.
The dimension of $\mathbf{L}$ is $|V| \times |V|$ where $|V|$ is the number 
of spectral elements in the mesh.  The unweighted Laplacian matrix $\mathbf{L} =
\left(l_{ij}\right)$ is defined as follows:
\begin{equation}
  \label{eqn:unweighted_laplacian}
  l_{ij} = \left\{
                   \begin{array}{ll}
                     -1 & \mbox{if } (v_i, v_j) \in E \\
                     deg(v_i) & \mbox{if } i = j \\
                     0 & \mbox{otherwise}
                   \end{array}
    \right.
\end{equation}
Here, $deg(v_i)$ is the degree of vertex $v_i$ in the graph.
In other terms, it is the number of neighbors of corresponding element $e_i$ in
the mesh.
Definition~(\ref{eqn:unweighted_laplacian}) is for the unweighted Laplacian since
connections between elements are not weighted, i.e., each connection between two
elements shows up in the Laplacian matrix as a $-1$ irrespective of whether they
share a vertex, edge, or face.
In other words, we have a constant (unit) weight function in this case.

In SEM/FEM meshes, it makes sense to put more weight to the latter connection since
if those two elements end up on different processors, the communication volume between the
processors is higher than if the elements shared just a single vertex.
Hence, we use the following weighted form of the Laplacian which tends to give
better partitions in practice:
\begin{equation}
  \label{eqn:wieghted_laplacian}
  l_{ij} = \left\{
                   \begin{array}{ll}
                     -\omega_{ij} & \mbox{if } (v_i, v_j) \in E \\
                     \sum_{i \ne j}{\omega_{ij}} & \mbox{if } i = j \\
                     0 & \mbox{otherwise}
                   \end{array}
    \right.
\end{equation}
where $\omega_{ij}$ is the weight associated with the edge $(v_i, v_j)$.
We define the weight function $\omega: E \rightarrow \mathbb{R}_{>0}$ associated
with the graph $G(V,E)$ to be the number of vertices shared between elements
$(e_i, e_j)$ corresponding to $(v_i, v_j)$ in $G(V,E)$.
For example, the weights for a 3D hex mesh would be 1, 2, or 4, depending
on whether the connection is by vertex, edge, or face.

The Laplacian matrix can be written as $\mathbf{L}=\mathbf{D}-\mathbf{A}$,
where $\mathbf{D}$ is the degree matrix and $\mathbf{A}$ is the adjacency
matrix. $\mathbf{D}$ is a diagonal matrix that contains the degree of each
vertex for the unweighted Laplacian and the sum of weights for the weighted
Laplacian.
For the unweighted Laplacian, ${ij}^{th}$ entry of $\mathbf{A}$ is $1$ if there
is an edge between the elements $e_i$ and $e_j$ and zero otherwise.  For the
weighted Laplacian, the value of these non-zero elements of $\mathbf{A}$ is
equal to the weight.

The Laplacian matrix reveals interesting spectral properties of the
corresponding graph due to its close relation to the Laplacian
operator in partial differential equations (PDEs)~\cite{barnard1994fast}.
Spectral properties of the Laplacian matrix have been studied by many authors,
notably by Fiedler in~\cite{fiedler1973algebraic, fiedler1975property}.
It is easy to see that the Laplacian matrix is positive semi-definite and hence
all the eigenvalues are greater than or equal to zero.
The smallest eigenvalue of the Laplacian matrix is zero with a corresponding
eigenvector of all ones ($\underline{1}$).
This is because the row-sum of both weighted and unweighted Laplacians is
zero.
Let the eigenvalues of $\mathbf{L}$ be ordered as follows:
\begin{equation}
  \lambda_1 = 0 < \lambda_2 \le \lambda_3 \ldots \le \lambda_k
\end{equation}
where $k = |V|$. The multiplicity of $\lambda_1$ is equal to the number of
connected components of the graph.
Since all the SEM/FEM meshes are connected, the multiplicity of $\lambda_1$ is 1.
$\lambda_2$, known as the algebraic connectivity, is related to the edge and vertex
connectivities~\cite{pothen1990partitioning}.
$\lambda_2$ is a good measure of how well the graph is connected~\cite{de2007old}.

Let the eigenvector associated with $\lambda_2$ be $\underline{y}_2$.
This is known as the \textit{Fiedler vector}.  Fiedler studied the
partitions generated by the sorted components of $\underline{y}_2$ and proved
that vertices of the graph $G$ can be partitioned into connected subgraphs
based on the order of component values.
Based on this result, a graph bisection is obtained by first sorting the
components of $\underline{y}_2$ and then assigning the elements corresponding
to the first half of the components to the first half of processors and the
rest of the elments to other half of processors.
$\underline{y}_2$ can be found by a suitable numerical method for solving the
following eigenvalue problem and selecting the eigenvector associated with
the second smallest eigenvalue:
\begin{equation}
  \label{eqn:eigen0}
  \mathbf{L}\underline{x} = \lambda\underline{x}
\end{equation}
Since we know that the eigenvector associated with the smallest eigenvalue is all
ones ($\underline{1}$), we can restate (\ref{eqn:eigen0}) as finding the eigenvector
corresponding to the smallest eigenvalue which satisfies the following conditions:
\begin{equation}
  \label{eqn:eigen1}
  \begin{array}{l}
    \mathbf{L}\underline{x} = \lambda \underline{x} \\
    \underline{1}^{T}\underline{x} = 0
  \end{array}
\end{equation}
We could use inverse power iteration to solve~(\ref{eqn:eigen1}) with an initial vector
orthogonalized with respect to $\underline{1}$.
We can use a preconditioned conjugate gradient iteration to solve the Laplacian system
at each iteration of inverse power iteration with LAMG as the preconditioner~\cite{livne2012lean}.
Inverse power iteration is expensive since it involves solving a linear system
at each iteration.
An alternative method is to use Lanczos, which can be used to find the extreme
eigenvalues and associated eigenvectors.
We have implemented both of these methods, with multigrid-preconditioned
conjugate gradient iteration to solve Laplacian system in inverse power
iteration.  

Once $\underline{y}_2$ is found, the graph is bisected as described
above.  Partitioning continues recursively in each half till each partition is
entirely contained in a single processor.
Algorithm~\ref{algo:rsb} outlines the RSB algorithm.
\begin{algorithm}
\caption{Recursive Spectral Bisection (RSB)} \label{algo:rsb}
\begin{algorithmic}[1]
  \State Find $y_2$ using Lanczos or Inverse Power Iteration
  \State Sort mesh elements according to $y_2$
  \State Assign $1^{st}$ half of elements to $1^{st}$ half of processors
  \State Assign $2^{nd}$ half of elements to $2^{nd}$ half of processors
  \State Recurse on two halves till they are entirely contained in a single processor
\end{algorithmic}
\end{algorithm}
Going down the recursive tree in Algorithm~\ref{algo:rsb}, if we run into a
situation where the number of processors is not divisible by two, we can create
two partitions with number of processors in each of them differ by $1$.  In
general, If $P$ is the number of processors at any given level, we can create
two partitions with sizes $\lfloor \frac{P}{2} \rfloor$ and $\lceil \frac{P}{2}
\rceil$.  Elements in the mesh will also be distributed between two partitions
proportional to the number of processors in each partition.

\section{Evaluating the Laplacian}\label{sec:laplacian}

Both Lanczos and inverse power iteration require performing the action of the
Laplacian on an input vector.  The most straightforward way to do this is
constructing the Laplacian matrix explicitly and store it in a sparse matrix
format like Compressed Sparse Row (CSR), with rows distributed among different
processors.  Then a sparse matrix vector product can be performed simply by
gathering required vector components in each processor.
The other option is to \textit{partially assemble Laplacian} simply by keeping
track which values need to be summed up while performing the action of Laplacian
on the vector. We call this the \textit{gather-scatter} approach.

\begin{figure*}
  \centering
{\setlength{\unitlength}{1.0in} \begin{picture}(6.500,3.00)(0,-.0)
  \put(0.00,0.00){\includegraphics[width=2.6in]{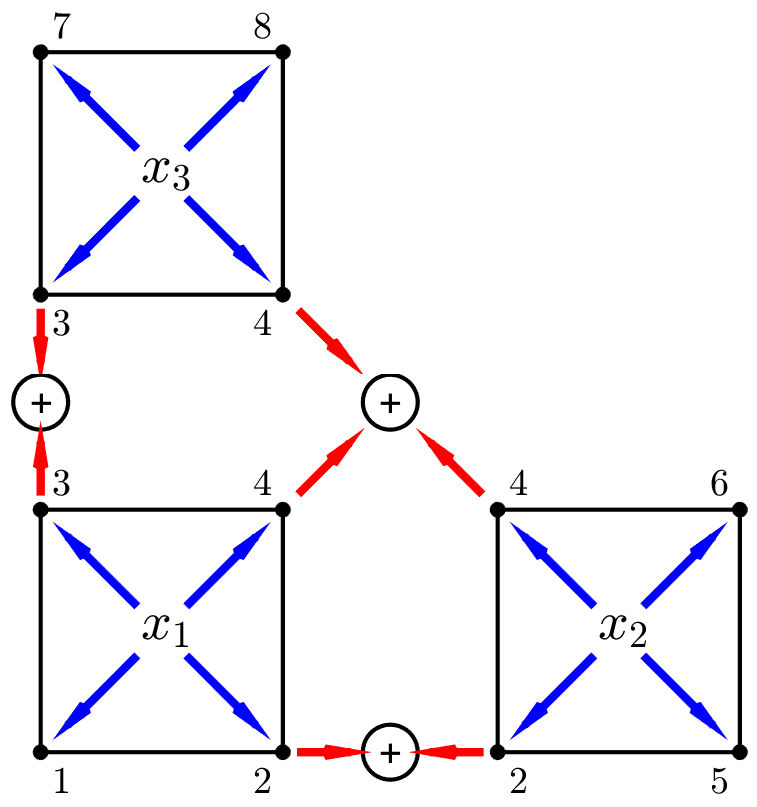}}
  \put(3.60,0.00){\includegraphics[width=2.6in]{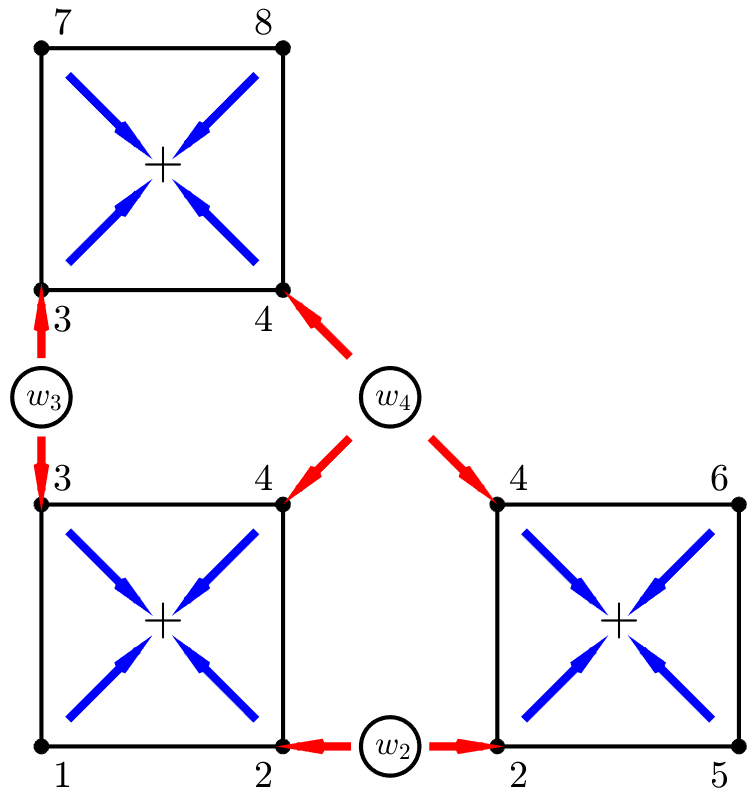}}
\end{picture}}
\caption{Application of (left) $Q^TP$ and (right) $P^TQ$.
Blue arrows indicate the application of $P$ and $P^T$, which is local to each
element (and hence to each processor), while red arrows indicate the
application of $Q$ and $Q^T$, which require inter-processor communication
in cases where adjacent elements are not on the same processor.
At the end, one has $\uy = P^T Q \uw = P^T Q Q^T P \ux$.
$Q$ and $P$ are copy operations, while $Q^T$ and $P^T$ involve summation.
Global vertex numbers are indicated near the $v=4$ vertices of each
element.
\label{fig:gs}}
\end{figure*}

In SEM/FEM applications, the {\em gather-scatter} operation is a central part
of the matrix-assembly and load-vector assembly processes.  In the PDE context,
for a given mesh with $m$ elements, one has $m_L = m \times v$ local vertices
that, for linear elements, correspond to local degrees-of-freedom.  These local
degrees-of-freedom allow for easy, element-based, evaluation of derivatives and
integrals.  However, for continuous finite element approximations, the true (or
global) degrees-of-freedom are associated with the shared nodes---local nodal
values are simply copies of these global values.  Suppose there are $m_u$ of
the unique global nodal values, $\uu=\{u_j\}$, enumerated $j=1,\dots,m_u$.
Then a map (or copy) from the global to representation can nominally be
effected by the matrix-vector product, 
\begin{eqnarray} \label{eq:qu}
\uu_L &=& Q \uu,
\end{eqnarray} 
where $Q$ is a sparse $m_L \times m_u$ Boolean matrix that has a single nonzero
in each row.  Each column of $Q$ corresponds to a single entry in $\uu$ and the
number of nonzeros in that column is equal to the number of elements that share
the global id associated with that column.  We refer to (\ref{eq:qu}) as the
{\em scatter} operation---data is scattered to the elements from the global
degrees-of-freedom.  Equally important in SEM/FEM applications is the {\em
gather} operation, $\uw = Q^T \uu_L$.  In this case, local values are {\em
summed} to their global counterparts.  

Figure \ref{fig:gs} illustrates the application of $Q^T$ (left) and $Q$ (right)
in red.  In the gather operation ($Q^T$) values on local nodes that are shared
by more than one element are mapped, with summation, to their global counterparts
(denoted by $\oplus$ in the left figure).  If this operation is followed by
a scatter ($Q$), these summed values ($w_j$ in the right figure) are copied
back to the originating local vertices.  

For the graph Laplacian on the dual of the vertex graph (i.e., on the graph
associated with element centroids), we need to map one input value, $x_e$, 
from each element to one output value, $y_e$, on each element.  This operation
can be expressed with a local copy matrix, $P$ that first maps $x_e$ for
element $e$ to its $v$ vertices.  At that point, one applies the nonlocal
gather-scatter, $QQ^T$, followed by $P^T$ which sums the $v$ values back to
$y_e$.  $Q^TP$ is illustrated on the left and $P^TQ$ on the right of Fig.
\ref{fig:gs}.

For the dual graph, the weighted adjacency matrix can be represented as
\[
  A_w = P^T QQ^T P.
\]
The weighted degree matrix $D_w=$diag($d_e$), with $\ud = A_w\underline{1}$
representing the vector containing the row-sums of of $A_w$.
From these two matrices we have $\mathbf{L}=D_w - A_w$ as the graph Laplacian.
To understand how this approach leads to the weighted Laplacian, consider a
case with $x_1$=1 and $x_2=x_3=0$.  We see that $\uz= A_w \ux$ will generate
$z_2=2$ because the value at $x_1$ has two paths to reach element 2.
Singletons, such as global vertex 1, will make no contribution to 
$\uy=\mathbf{L} \ux$ because their contribution is cancelled out by the
difference $D_w - A_w$.

The advantage of the gather-scatter approach is that it can be implemented with
minimal setup cost, which is particularly important given that, at the
partitioning stage, the problem setup has yet to be well defined on the
processors.   The simulation starts with what is effectively a random
distribution of elements assigned to each processor, where nothing is known
about how the graph is mapped to the processors.   (As discussed in Section
\ref{sec:results}, we have found it advantageous to use an initial pass of RCB
to at least organize the data into contiguous subsets in order to minimize
gather-scatter communication overhead during Lanczos or inverse iteration.) To
set up the $QQ^T$ communication, we use the gather-scatter library, {\em gslib}
\cite{gslib}, which involves two calls, {\em gs\_setup}, and {\em gs\_op}.  In
the setup phase, each processor provides an $m_p \times v$ list of global
integer (long) pointers corresponding to the global id of each vertex on each
of the $m_p$ elements on rank $p$.  In a discovery phase, {\em gs\_setup}
discerns which ranks have commonly shared vertices, which vertices are multiply
represented within a rank, and which are singletons.
Subsequently, one calls {\em gs\_op} with a list of
$m_p \times v$ doubles (or floats), to which $QQ^T$ is applied.
The setup phase executes in $O(\log P)$ time and typically runs
multiple trials of {\em gs\_op} using either a pairwise exchange, a generalized
all-to-all (based on the crystal-router of \cite{fox88}), or an all-reduce,
choosing whichever is the fastest.  Users can also prescribe a prefered
communication algorithm if they wish to reduce the setup overhead.

We illustrate the {\em gs\_setup} call for the the example of Fig. \ref{fig:gs}
under the assumption that elements 1 and 2 are on rank 0 while element 3
is on rank 1.  On respective ranks, the setup differs only in the
enumeration of the global ids,
\begin{eqnarray*}
&&\mbox{rank 0:  \hspace{.1in}
   global\_id=[ 1 2 3 4 2 5 4 6 ]; \hspace*{.05in}
   $m_L$ = 8; } \\
&&\mbox{rank 1:    \hspace{.1in}
   global\_id=[ 3 4 7 8 ]; \hspace*{.530in}
   $m_L$ = 4; }
\end{eqnarray*}
Thus defined, respective setup and execution of $QQ^T$ would be
expressed on each processor by the commands
\begin{eqnarray*}
&&\mbox{
gs\_handle = gs\_setup(global\_num,$m_L$);} \\
&&\mbox{
gs\_err \hspace*{.171in} = gs\_op(gs\_handle,w); } 
\end{eqnarray*}
The {\em gs\_op} call in this case generates the result $\uw:=QQ^T\uw$.
(We remark that {\em gslib} has scaled to millions of ranks on the IBM 
BG/Qs, Mira and Sequoia, with setup times typically measured in fractions
of a second.  For 1 billion global vertices, the setup time on 1048576
ranks of Mira was 1.24 seconds, including 20 trial runs of {\em gs\_op}.
Setup times for current examples on Summit are provided in Section 
\ref{sec:results}.)

Evaluating the unweighted Laplacian is more complex than the weighted
Laplacian since we must avoid adding contributions from the same neighbor
element multiple times.
In other words, connection between two elements should be counted only once
irrespective how many vertices they share.
The unique set of neighbors of an element can be found by a simple counting
principle.
In order to do so, we need to think of the connectivity between two elements
in terms of vertices, edges and faces.

When two elements are connected in a SEM mesh, they are either connected by
a vertex, edge or a face in 3D.
In 2D, we only have vertices and edges.
The weight of the connection is the total number of vertices they share.
A neighbor which shares an edge is counted twice and a neighbor which shares a
face is counted four times in the weighted Laplacian.
We can adjust the gather sum for edges by removing contributions once for each
shared edge.
An undesirable side effect of this subtraction is that the contributions from
face neighbors are eliminated since they have 4 vertices and 4 edges in common.
So we need to add back contributions if there is a shared face.
If we can number edges and faces of the mesh uniquely, we can setup the
gather-scatter such that contributions from each neighbor element is counted
exactly once.
It turns out that it is very easy and fast to do this numbering as we
have a global numbering for vertices already available.


\section{Lanczos Iteration}\label{sec:lanczos}

The Lanczos algorithm is preferred for finding eigenvalues of real symmetric matrices
since the number of floating point operations per iteration is smaller than other
equivalent methods like Arnoldi~\cite{wu2000thick}. We implemented Lanczos
with restarts to find $\underline{y}_2$.

Suppose Lanczos algorithm takes $j$ number of iterations before it reaches one
of terminal conditions (maximum number of iterations or the tolerance).
Lanczos algorithm produces as output a series of vectors $\underline{q}_i$ and scalars
$\alpha_i$ and $\beta_i$ where $ 1 \le i \le j$.
The vectors $\underline{q}_i$ are known as \textit{Lanczos vectors} and are
orthonormal.
Let $\mathbf{S}$ be the $n \times j$ matrix which has the Lanczos vectors as the
columns.
\[\mathbf{S} = \left[\begin{array}{cccc}
  \mid & \mid & & \mid \\
  \underline{q_1} & \underline{q_2} & \dots & \underline{q_j} \\
  \mid & \mid & & \mid \\
\end{array} \right] \]
The scalar series $\alpha_j$ form the diagonal of a symmetric tri-diagonal
matrix $\mathbf{T}$ while $\beta_j$ form the subdiagonal (and superdiagonal) of
$\mathbf{T}$.

\[\mathbf{T} = \left[ \begin{array}{cccc}
  \alpha_1 & \beta_2 & & \\
  \beta_2 & \alpha_2 & \beta_3 & \\
  & \ddots & \ddots & \ddots \\
  & & \beta_j & \alpha_j \end{array} \right] \]

The eigenvalues of $\mathbf{T}$ are known as Ritz values. If $\underline{t}_i$, $1\le i \le j$
are the eigenvectors of $\mathbf{T}$, the vectors $y_i = \mathbf{S} \underline{t_i}$ are
called the Ritz vectors.
Ritz values and vectors are the approximations for the eigenvalues and eigenvectors
of $\mathbf{L}$.
We can find an approximation to $\underline{y_2}$ by finding the eigenvector corresponding
to minimum eigenvalue in $\mathbf{T}$ and then multiplying it by $\mathbf{S}$.


\section{Inverse Iteration}\label{sec:inverse}

\begin{algorithm}
\caption{Inverse Iteration} \label{algo:inverse}
\begin{algorithmic}[1]
  \State Initialize $\underline{b}$ with random values
  \For{$i \gets 1$ to $j$}
    \State Orthogonalize $\underline{b}$ w.r.t $\underline{1}$
    \State $\underline{b} = \underline{b}/\lVert\underline{b}\rVert$
    \State Solve $L\underline{y_2} = \underline{b}$
    \State $\underline{b} = \underline{y_2}$
  \EndFor
\end{algorithmic}
\end{algorithm}


We implemented inverse iteration (Algorithm~\ref{algo:inverse}) using a
preconditioned flexible conjugate gradient (flexcg) method, augmented with full
projection to solve for Fiedler vector once we reached larger problem sizes for
which the Lanczos didn't converge.
A key point of our flexcg implementation is that the initial search direction is
{\em not} preconditioned.  This choice is motivated by the fact that we
are solving for $\uy_2$, which is an eigenvector of $\bL$ and that the
preceding iterate is used as the right-hand side of $\bL \ux = \ub$.
As $\ub \longrightarrow \uy_2$, the Krylov subspace (in $\bL$, but not
in $M^{-1} \bL$) will be invariant and the modified flexcg will return
in a single iteration, at which point we stop the outer inverse-iteration.
We used an aggregation-based algebraic multigrid (AMG) preconditioner inspired
by the Lean Algebraic Multigrid~\cite{livne2012lean} method.
Pseudocode for the multigrid preconditioner is shown in Algorithm~\ref{algo:vcycle}.

\begin{algorithm}
\caption{V-Cycle} \label{algo:vcycle}
  \hspace*{\algorithmicindent} \textbf{Input:} $\sigma, n_{smooth}, i, \underline{r}, L$ \\
  \hspace*{\algorithmicindent} \textbf{Output:} $\underline{u}$
  \begin{algorithmic}[1]
      \State $D = 1.0/diag(L)$
      \State $\underline{u} = \sigma\,D\underline{r}$
      \State $r = r - L\underline{u}$
      \For{$s \gets 1$ to $n_{smooth}$}
        \State $\underline{u} = \underline{u} + \sigma D \underline{r}$
        \State $\underline{r} = \underline{r} - L \sigma D \underline{r}$
      \EndFor
      \State $L_c = J_{i}^{i+1} L J_{i+1}^{i}$
      \State $\underline{r_c}= J_{i}^{i+1} \underline{r}$
      \State $\underline{e_c} = $ V-Cycle$(\sigma, n_{smooth}, i, \underline{r_c}, L_c)$
      \State $\underline{u} = \underline{u} + J_{i+1}^{i} \underline{e_c}$
      \For{$s \gets 1$ to $n_{smooth}$}
        \State $\underline{r} = \underline{r} - L \underline{u}$
        \State $\underline{u} = \underline{u} + \sigma D \underline{r}$
      \EndFor
  \end{algorithmic}
\end{algorithm}

The finest level of multigrid, level $l_{0}$, uses the Laplacian corresponding to
the actual element mesh, which is evaluated using either the CSR or
gather-scatter representation.
For coarser levels ($l_{1}, l_{2}, \hdots$), the gather-scatter based Laplacian
evaluation is not straightforward to setup as we don't have a global vertex
numbering readily available once the elements are coarsened.
 For those levels, we forget about the geometric aspect of coarsening
and view it as simply applying a restriction operator $J_{i}^{i+1}$
going from level $i$ to $i+1$ and its transpose $J_{i+1}^{i}$ to generate
$L_{i+1}=J_i^{i+1} L_{i} J_{i+1}^i$, where $J_i^{i+1}$ is a Boolean
matrix corresponding to piecewise-constant prolongation.
It is easy to see that this preserves the qualities of the Laplacian.

We bootstrap the prolongation operator from an RCB ordering of the mesh elements.
Before setting up the multigrid $V$-cycle, we partition the mesh using RCB.
Denote the coarsest level as $l_N$, which comprises all elements.
The next level comprises the two sets arising from the initial coordinate bisection,
such that the prolongation matrix is $J_{N}^{N-1} = [\, 1 \; 1 \, ]^T$.  If $I_2$ is the
rank-2 identity matrix, then $J_i^{i-1} = I_2 \otimes J_{i+1}^i$, up to the finest
level, where the number of entries in $L_0$ is typically not a power of 2.
Note that $L_{1} = J_{0}^{1}L_{0}J_{1}^{0}$ will be approximately half as large
as $L_0$ because aggregating element pairs at level $l_0$ is equal to
condensing (i.e., summing) the rows and columns of $L_0$ corresponding to those
pairs.

Our aggregation strategy is local to a processor till we reach a single row per
processor and we keep number of processors constant till we reach that threshold.
Once we reach that threshold, we turn off the number of processors by half at
each level since the size of the Laplacian matrix is reduced by half.
This process is described in detail below.

\begin{enumerate}
  \item Assume that at the level $l$, we have a $n_l \times n_l$
    matrix and $P_l$ active processors.
    Also, Laplacian $L_l$ is represented using CSR format.
  \item If $n_l = P_l$, we set $P_{l+1}=P_l/2$, otherwise $P_{l+1}=P_l$.
    We will keep processors $p_1,\hdots p_{P_{l+1}}$ active in level $l+1$.
  \item Gather rows $1, \hdots n_l$ of level $l$,
    distributed over $P_l$ processors, to $P_{l+1} \leq P_l$ processors.
    (If $P_l=P_{l+1}$, no data movement takes place.)
  \item Collapse these rows locally on $P_{l+1}$ processors to get
    $n_{l+1}$ rows where $n_{l+1} \approx n_l/2$.
    (The orginal rows are preserved; the collapsed outputs
    are written to a new row id in level $l+1$.
  \item Repeat the above for columns as well.
  \item These steps produce a mapping of row and column ids from level $l$ to
    $l+1$ which can be used to setup $J_i^{i+1}$ and $J_{i+1}^{i}$.
%
%
\end{enumerate}

With the above steps, we generate $L_0, L_1, L_2, \hdots$ as CSR matrices.

\section{Results}\label{sec:results}

We tested {\em parRSB} as a mesh partitioner for NekRS\cite{fischer2021nekrs} -- a highly scalable
GPU-accelerated spectral element solver.
Tests were done on Summit and Frontier supercomputers at OLCF where {\em parRSB} was run on the CPU.
Running {\em parRSB} on CPU is not a bottleneck for NekRS since parRSB finished in less than a
minute during all our tests.
In the context of the SEM simulations where the setup time could easily take minutes due to various
preprocessing steps and the preconditioner setup, the time taken by {\em parRSB} is negligible.
In SEM production runs, total setup time is only a fraction of the time taken by the solver since the
actual simulation time could easily be hours or days.

We made sure that the partitions generated by {\em parRSB} are load balanced as much as possible
with the maximum load imbalanced being only a single element as mentioned in Equation~\ref{eq:partition}.
To evaluate the partition quality, we choose number of neighbors and the average communication volume
per neighbor (i.e., total outgoing message size from a processor divided by the number of neighbors).
As pointed out in Section~\ref{sec:intro}, these two metrics are responsible for the communication.

We used two pebble bed meshes with 13 million and 99 million spectral elements for comparing Lanczos and
inverse iteration on Summit supercomputer.
For both Lanczos and inverse iteration, we evaluated Laplacian operator using the gather-scatter approach
listed in Section~\ref{sec:laplacian}.
Gather-scatter setup times for the 99 million element mesh was about 0.15s for 115 billion vertices
using 37800 processors.
For the 13 million mesh, it took about 0.037s for 15 billion vertices with 11340 processors.
These setup times are a fraction of the time taken to run either Lanczos or inverse iteration.

\begin{table*}[htb]
  \centering
  \begin{tabular}{|r|r|r|r|r|}
    \hline
    $P$  & Lanczos Time (s) & RCB + Lanczos Time(s) & Max Neighbors & Average Neighbors \\ \hline
    4872 & 45.5             &   24.8                & 25            & 16.0              \\
    6468 & 36.8             &   20.4                & 26            & 16.5              \\
    8106 & 30.5             &   17.2                & 27            & 17.0              \\
    9744 & 30.6             &   16.6                & 30            & 16.8              \\
    11340& 29.1             &   16.4                & 27            & 17.0              \\ \hline
  \end{tabular}
  \caption{Partition time and number of neighbors for 13 Million elements mesh using Lanczos.}
  \label{table:13m_lanczos}
\end{table*}

\begin{table*}[t]
  \centering
  \begin{tabular}{|r|r|r|r|}
    \hline
    $P$  & Inverse Time (s) & Max Neighbors & Average Neighbors \\ \hline
    4872 & 32.8             & 28            & 16.2              \\
    6468 & 30.0             & 27            & 16.6              \\
    8106 & 28.8             & 35            & 17.2              \\
    9744 & 30.0             & 32            & 17.0              \\
    11340& 28.5             & 29            & 17.1              \\ \hline
  \end{tabular}
  \caption{Partition time and number of neighbors for 13 Million elements using preconditioned inverse iteration.}
  \label{table:13m_inverse}
\end{table*}

\begin{table*}[htb]
  \centering
  \begin{tabular}{|r|r|r|r|}
    \hline
    $P$   & RCB + Lanczos Time (s) & Max Neighbors & Average Neighbors \\ \hline
    16212 & 67.2                   & 31            & 17.2              \\
    21588 & 57.7                   & 27            & 16.3              \\
    27006 & 50.4                   & 26            & 16.6              \\
    37800 & 43.6                   & 29            & 16.8              \\ \hline
  \end{tabular}
  \caption{Partition time and number of neighbors for 99 Million elements mesh using Lanczos.}
  \label{table:99m_lanczos}
\end{table*}


Timing data and number of neighbors for the 13 million element mesh is shown in Table~\ref{table:13m_lanczos}
as the number of processors is increased from 4872 to 11340.
First column of the table is the number of processors and second column is the time taken to finish
the partitioning with Lanczos.
We found that running a geometric partitioning scheme like RCB or RIB before RSB can significantly
reduce time taken by the RSB algorithm.
Lanczos runtime reduced approximately by $2\times$ when RCB is used as a pre-partitioner.
This gain is more pronounced at the higher levels of the bisection tree where the global partition
sizes are larger.
As we go down the bisection tree running RCB or RIB as a pre-partitioner has diminishing returns.
Third column is the time taken to finish the partitioning with RCB as a pre-partitioner for Lanczos.
Fourth and fifth columns show maximum number of neighbors and the average neighbors respectively.

Table~\ref{table:13m_inverse} shows the same information for preconditioned inverse iteration.
These results are comparable to Lanczos (in Table~\ref{table:13m_lanczos}) except when it comes to
execution time.
Main culprit for slowing down inverse iteration is the AMG preconditioner setup, which has to be done
at each level in RSB tree, not the inverse iteration itself.
For the mesh under consideration, for $P = 4872$, both Lanczos and inverse iteration spent about $2s$
finding the Fiedler vector for the first cut.
Inverse iteration took about 6 iterations to find the cut compared to Lanczos which reached the maximum
number of restarts (50) permitted in our implementation.
Table~\ref{table:99m_lanczos} list results for 99 million element pebble-bed mesh.
We can see that the average number of neighbors stay more or less the same as the number of processors
are increased.
Run time for Lanczos for the 99 million element mesh is higher than the 13 Million element mesh due to
the larger local problem sizes.

\begin{table*}[htb]
  \centering
  \begin{tabular}{|r|r|r|r|r|}
    \hline
    $P$   & RCB + Lanczos Time (s) & Max Neighbors & Average Neighbors & Average Message Size\\ \hline
     8    & 0.1                    & 7             & 7.0               &  8580\\
     512  & 1.5                    & 26            & 19.8              &  5600\\
     1024 & 6.0                    & 23            & 14.0              &  9890\\
     4096 & 6.0                    & 26            & 23.0              &  5130\\
     8192 & 20.0                   & 24            & 15.6              &  8680\\
     16384& 20.0                   & 25            & 15.8              &  9092\\
     32768& 20.0                   & 26            & 24.0              &  4890\\
     65536& 41.0                   & 27            & 16.0              &  8427\\ \hline
  \end{tabular}
  \caption{Partition time and quality for a weak scaling study with $E/P\approx8000$.}
  \label{table:weak_frontier}
\end{table*}

Table~\ref{table:weak_frontier} shows results for a weak scaling study on the Frontier supercomputer using
cube meshes as the number of processors is increased from $8$ to $65536$ while the number of elements per
processor is kept constant at $\approx8000$.
We can see that {\em parRSB} generates partitions with maximum and average number of neighbors in the expected
range (26) for a SEM mesh.
What is more interesting is the average message size reported in the last column.
Given that the $m_2$ for Frontier is about 5000 (\textbf{TODO: Cite ping pong tests}), based on the
numbers reported for the average message size, we can see that the most of these runs fall on the regime
where communication volume is the dominant factor in communication time.


\section{Future Work}\label{sec:future}

It is not uncommon in graph Laplacians to have eigenvalues with
multiplicty greater than 1.  For example a checkerboard would have
a Fiedler-vector pair that shares the same minimum nontrivial eigenvector.
In fact, the space is much larger, because things only need to be
{\em topologically} a checkerboard to have this property.  So, a
rectangular mesh with the same number of elements in $x$ and $y$
would have a matched pair of Fiedler vectors, and {\em any}
linear combination of these eigenvectors is a valid Fiedler vector
in the RSB algorithm.  Unfortunately, the quality of the cut is
not the same for all vectors.  Cutting along the vertical or horizontal
midline of an $N\times N$ graph will expose $N$ faces; cutting on a
45$^o$ angle would expose $\approx 2N$ faces, which is suboptimal.
Conceivably, one could explore the one dimensional parameter space
spanned by $\theta\uv_1 + (1-\theta) \uv_2$, which constitutes all
linear combinations of the eigenvector pair to try to find the min-cut
solution.

Unfortunately, neither standard Lanczos nor inverse iteration will reveal a
second eigenvector in this pair---what emerges is merely the same linear
combination of those eigenvectors that was in the initial seed vector for the
Lanczos/inverse iteration.  To find the second eigenvector, one would need
to run a block, or subspace, iteration comprising two vectors that are
orthogonalized on each iteration.  Fortunately, the overhead for performing
such a block iteration is unlikely to be large given that these algorithms
are dominated by relatively sparse (i.e., latency-limited) communication.
Moreover, {\em gslib} already supports vector-based communication, that is,
multiple values at each node of the graph.  In cases limited by interprocessor
latency, the cost of communicating extra values at each node is close to nil.
In addition, a block algorithm would amortize the setup overhead of the graph
and of the aggregation-based AMG for inverse iteration.   We believe that
such an algorithm might be of value in generating higher-quality RSB-based
cuts and are pursuing these ideas with that goal in mind.
There are multiple ways to reduce setup cost of the AMG preconditioner.

%

$10^8$ element meshes are just a start and future exascale computers will be
running billion element meshes.
For SEM/FEM simulations, CPU based supercomputers like Mira at Argonne Leadership
Computing Facility used to run efficiently with 2000 grid pointers per rank.
With GPGPUS, Summit at OLCF usually requires about 2 million grid points per rank
in order to run at an acceptable efficiency.
We expect this trend to continue and the local probem size in exascale systems
most likely be larger than that of Summit.
For a larger local problem size, it will take longer for parRSB to just run on CPUs.
So porting \textit{parRSB} to use accelerators is in our roadmap.

\bibliographystyle{unsrt}
\bibliography{all}

\end{document}